\documentclass[runningheads]{llncs}
\usepackage{graphicx}
\usepackage{url}
\usepackage{array}
\begin{document}
\title{Creativity and Visual Communication from Machine to Musician: \\Sharing a Score through a Robotic Camera}
\titlerunning{Creativity and Visual Communication from Machine to Musician}
%
\author{Ross Greer\inst{1, 2}\orcidID{0000-0001-8595-0379} \and
Laura Fleig\inst{1}\orcidID{0009-0003-0982-2641} \and
Shlomo Dubnov\inst{1}\orcidID{0000-0003-0222-1125}}
\authorrunning{Greer et al.}
%
\institute{University of California San Diego, La Jolla, CA, USA \and
University of California Merced, Merced, CA, USA
}

%
%
\maketitle              
\begin{abstract}
This paper explores the integration of visual communication and musical interaction by implementing a robotic camera within a ``Guided Harmony" musical game. We aim to examine co-creative behaviors between human musicians and robotic systems. Our research explores existing methodologies like improvisational game pieces and extends these concepts to include robotic participation using a PTZ camera. The robotic system interprets and responds to nonverbal cues from musicians, creating a collaborative and adaptive musical experience. This initial case study underscores the importance of intuitive visual communication channels. We also propose future research directions, including parameters for refining the visual cue toolkit and data collection methods to understand human-machine co-creativity further. Our findings contribute to the broader understanding of machine intelligence in augmenting human creativity, particularly in musical settings.

\keywords{Human-robot interaction  \and human-machine co-creativity \and visual communication.}
\end{abstract}

\section{Introduction}  

Group music-making relies on a rich exchange of information between musicians, requiring multiple channels for sending and receiving messages, and agreed-upon conventions for which messages (or, more often, whose messages) will be given priority and attention to form a cohesive musical output. These messages come in a variety of forms: conversations before a rehearsal, eye contact and head nods during performance, collective breathing and motion at the onset of a phrase, adjustments to pitch to match what is heard from neighbors, and more. This communication, verbal or otherwise, contains two components: (1) an encoder who shares information with a predetermined code, and (2) a decoder who observes and responds. Successful communication, then, means that the encoder’s intention is known to both the encoder and the decoder. The extent to which the sender's intent matches the receiver’s perceived intent can be defined as the functional achievement of the communicational act, i.e., how successful it was \cite{juslin1997emotional}. The act of creating music with one or more others can be thought of as a kind of conversation. Consequently, musicians work with various verbal, nonverbal, and musical cues that assist them in transmitting information to, i.e., communicating with, their fellow musicians. Generally, a musical cue is “any verbal, visual, or auditory signal that directly influences the nature and direction of the performance” with the aim to “time the aesthetic tone” and “give a desired quality of sound” \cite{gordy1999duo}. While much of musical communication can occur over audio channels, nonverbal modes of communication also lead to rich musical expression. Physical gestures are used to indicate attacks and releases; eye contact is used to draw attention and importance \cite{greer2021restoring}; posture changes convey prominence for balance. Especially in ensemble settings, the conductor exercises musical communication in this strictly visual manner. 

Building on this understanding of musical communication, we explore the emerging paradigm of co-creative human-machine interaction, which extends the notions of control of generative AI to communication for mutual understanding of the goals, intent, and the contextual framework in which the human and machine operate. When multiple parties collaborate creatively, individual agents’ ideas build upon each other. Unlike simply dividing up tasks, co-creativity tends to lead to more inventive solutions, embodying the principle that the whole is greater than the sum of its parts \cite{davis2013human}. Specifically, we investigate how to facilitate co-creative behaviors between humans and machines through multimodal interactions. In our musical setting, we interpret visual cues with contextual meta-score data for the creation of more sophisticated and context-aware AI improvisation systems. 

In this paper, we introduce an initial case study by creating a musical game in such a framework. Toward continued research in human-machine co-creativity, we propose additional compositional games and extensions within this framework.

\section{Related Research}

How can a human or a robot communicate visually in a musical context? This section examines available visual modes of musical communication for humans and machines, aiming to integrate them effectively into our co-creative framework.

\subsection{Musical Games and Improvisation}

First, we turn our attention to human-to-human communication within musical games. Musical games have an established history as exploratory tools for understanding machine processes and creativity \cite{hedges1978dice}\cite{volchenkov2012markov}\cite{lin2015audio}\cite{dubnov2023deep}, and in this section, we explore how specific improvisational experiences such as conduction \cite{stanley2009butch}, Soundpainting \cite{Soundpainting}, Cobra \cite{van2013free}, and public conducting facilitate real-time, collaborative music-making.

One prominent figure in this area is Butch Morris, who developed the concept of ``conduction," a method of conducted improvisation/interpretation that allows a conductor to direct musicians using a ``predetermined vocabulary of bodily gestures" \cite{stanley2009butch}. Similarly, Walter Thompson's Soundpainting, a ``universal multidisciplinary live composing sign language," uses over 1500 gestures that let the composer communicate in real-time to the performers how the music should progress \cite{Soundpainting}. These systems highlight the importance of visual communication and real-time interaction to enable dialogue between conductor and musicians. 

John Zorn's ``Cobra," another notable example, is a game piece for musical improvisation involving a prompter who directs the ensemble using hand signals and cue cards. Performers can interrupt and influence the musical direction by making specific preset gestures to the prompter, creating a spontaneous collaborative musical experience \cite{van2013free}. These gestures include pointing at various body parts or making specific movements according to a score \cite{brackett2010some}. 

\begin{figure}[h]
    \centering
\includegraphics[width=\textwidth]{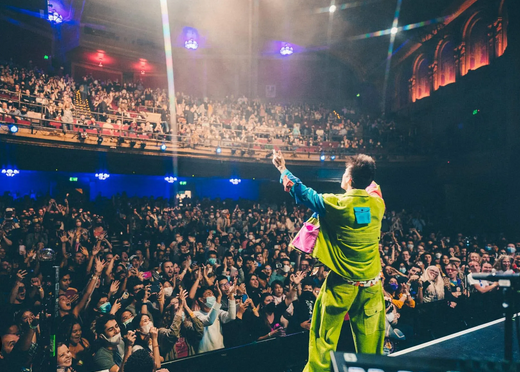}
    \caption{Jacob Collier leads a performance of ``Conducting the Audience". After providing initial pitches to two sections of the audience, he uses his hands to indicate to these sections to move their pitch up or down a step in a live performance.}
    \label{fig:jacob}
\end{figure}

In a more contemporary example, singer-songwriter Jacob Collier\footnote{\url{https://en.wikipedia.org/wiki/Jacob_Collier}} is well-known for public conducting (i.e., conducting his audience members) during live performances, engaging the audience in the creative process. He initializes different parts of the audience on different notes, then uses hand signals to direct their pitch up or down, and ultimately works through harmonic progressions. Acting as the creative leader, Collier guides the audience collectively toward his creative trajectory, creating a shared musical experience\footnote{https://www.youtube.com/watch?v=3KsF309XpJo}.

Building on these concepts, we recognize the importance of nonverbal communication between musical agents, typically through a set ``codebook" of visual cues. These can be arbitrary (such as touching one's ear like in Cobra) or more natural (such as Collier raising his hands to indicate a higher pitch). 

\subsection{Musical Robots}

Having examined the foundations of musical games, we now shift our focus to existing research on musical robots and their ability to interact with humans using visual cues.

One well-known example of a musical robot is Shimon, the robotic improvisational marimba player developed at Gil Weinberg's Georgia Tech Center for Music Technology. Shimon has mallets as arms and an expressive head with a digital video camera. It ``matches the human's playing style, tempo, and harmony in real time, while extending on the human's playing and contributing its own musical phrases and ideas” \cite{hoffman2010shimon}. Importantly, Shimon, while not exactly humanoid, can convey emotions. Head bobs visualize Shimon's internal beat, it makes eye contact to facilitate turn-taking, and it imitates human-like behavior through blinking and breathing \cite{hoffman2010shimon}. 

Most research into human-robot interaction focuses on how robots can understand human gestures through computer vision, with considerably less research into what kinds of robotic gestures are recognizable by humans. Visual intake of robotic action activates the same mirror neuron system in humans that fires when seeing biological action. This implies that it should be possible for robot mimicry of human movements to be accurately interpreted by a human observer \cite{cabibihan2012human}.

The specifics of how a robot might appropriately encode cues to send to a human musician greatly depend on the robot's physical structure. Given our research objectives, a PTZ (Pan-Tilt-Zoom) camera was selected as our robotic system for several reasons. While being a minimal system, it can mimic the act of looking at something, which is essential for interactive communication. Additionally, it can take in ``egocentric" camera input, functioning as another agent in the musical space. Finally, a PTZ camera provides a reasonable aesthetic for stage presence, making it a practical and visually appropriate choice. 

As of writing, there is no existing literature on imitating human head movements with a PTZ camera. However, combining the findings mentioned above leads to several proposed camera movements that might appropriately model corresponding human movements. Nodding and head shaking are easily representable through fast vertical and horizontal movements, respectively. Additionally, simulated eye contact can be established by having the camera turn directly to the human.

\subsection{Creativity}

Cognitive scientist Margaret Boden defines creativity as ``the ability to generate ideas/artifacts that are new, surprising, and valuable" \cite{boden2013creativity}. Computer scientist Jürgen Schmidhuber provides an alternative definition of the creative process as one that maximizes some reward for the creation of novel patterns, by which reinforcement learning methods may drive the continued creation of novel patterns.  When considering group creativity, however, the ability to generate comes not necessarily from individual agents but from their interactions. In this way, having an idea of a fixed reward or creative goal may not be sufficient since observation of the output of one agent may create new unconsidered possibilities for other creative agents in the process. This pattern can be observed on a large time scale, by which new research developments create further opportunities for creative and innovative responses; in this framework, we consider this phenomenon on the shortened and specific scale of group artistic creativity in music-making. As a recent and highly specific example, TikTok duets are a popular cultural media phenomenon, by which a soundbite from one user, deliberate or not, may be ``dueted" with, via direct audio overlay, sampling, remixing, or other means, creating a new set of creative output from initial creative source material\footnote{https://www.youtube.com/watch?v=ab15vGsXz9M}\footnote{https://youtube.com/shorts/kEHsT8jNIXw?si=C5-2OSXcdK5pZvzr}. Popular music jam sessions feature this same phenomenon, on a more ``real-time" production scale than social media affords. Similar to the structured improvisation ``games" described above, these collaborative creative acts build around one or more initial musical segments, sounds, or seeds, in complete or partial form, that are used as a leading material or inspiration to start a musical process, which then evolves as free-form or according to a pre-designed plan. 

In this research, we explore collaborative music-making with a centralized guidance agent, analogous to a conductor. We maintain the idea of adaptive agents (human or a generative machine bestowed with musical knowledge), whose creativity is coordinated through a centralized reward agent whose policy is defined through a partial musical score, but with a degree of freedom of moving along this score according to feedback received from the performing agents themselves. This model incorporates a bi-directional communication between the performing musicians and the conductor, whose goal is to coordinate an optimal joint performance within the constraints of the pre-defined score. Since the ``optimality" is defined not only in terms of following the musical score directives, but also incorporates instantaneous interaction, the model incorporates signals from visual interaction, gestures, and possibly future emotional and other meta-musical features. Moreover, the framework itself allows flexible or alternative score choices or musical trajectories, such as an open musical form or idea of open art in general \cite{eco1989open}.

\section{Case Study: ``Guided Harmony"}

Based on the above, we have designed a “guided harmony” musical game that utilizes a bidirectional codebook of visual cues to facilitate musician-to-machine and machine-to-musician communication. This game is intended to be an initial proof-of-concept for using a robotic camera for this, so we are limiting ourselves to the following modes of communication: 
\begin{itemize}
    \item The only human cue is a raised hand, which expresses discontentment with the current state. Since musical instruments put humans in poses that are not commonly trained on in pose detection systems, choosing a large enough cue for standard pose detection libraries to identify robustly and reliably is vital. (Incidentally, this highlights the need to create datasets of images of humans playing instruments.) Our prototype system uses MMPose Coco Wholebody, which has been used in applications which make decisions from modeled keypoints of human pose, including those of the hands and face \cite{rangesh2021autonomous} \cite{greer2023safe}.
    \item Additionally, this game limits machine movements to finding and centering on a specific musician (``eye contact"), nodding, and shaking its ``head". As a result, the joint output is human-only audio. 
\end{itemize}  

In our case study, the machine acts similarly to Collier's conducting, with a score guiding the performance. However, musicians retain some control, like in Cobra, where they can signal discontentment with the current musical state.

\begin{figure}[h]
    \centering
    \includegraphics[width=.4\textwidth]{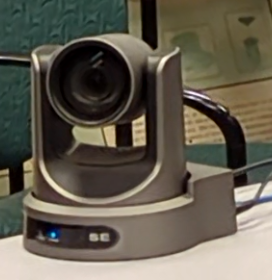} \\
    \includegraphics[width=.346\textwidth]{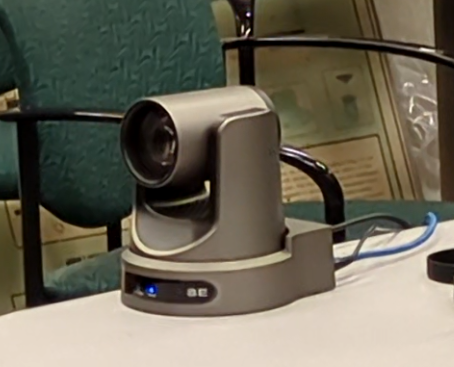}
    \includegraphics[width=.317\textwidth]{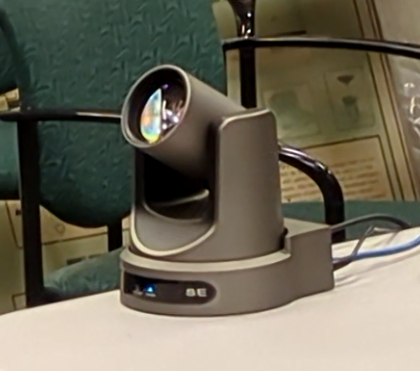}
    \includegraphics[width=.32\textwidth]{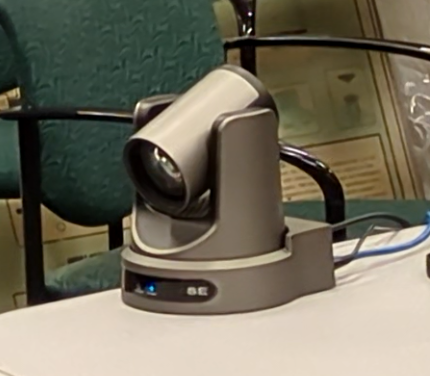}
    \caption{(Top) The ImproVision robot turns to make pseudo ``eye contact" with the violinist. After turning to indicate who ImproVision is communicating to, it will then provide an instruction through its motion. (Bottom) The process of providing an ensemble down-beat: the ImproVision robot first looks to ensemble center, lifts, then bows its head.}
    \label{fig:gestures}
\end{figure}

This guided harmony game has the concrete goal of having a machine guide a group of human musicians through a set score of sustained chords. This score is known to the machine and hidden from musicians (similarly to how Jacob Collier alone decides what chord progressions to move the audience through). The machine provides an initialization harmony to the (human) musicians, who sustain the sound. Musicians have only one communication channel to the machine, with only two states: ``content” or ``not content." The cue for ``not content” is a raised hand. When the machine receives a ``not content" cue (i.e., detects a raised hand) from any musician, it provides instructions to all musicians to adjust the harmony in a specific way. These are communicated through a two-part message – who should adjust, and how they should adjust. Currently, adjustment is constrained to the following: 
\begin{itemize}
    \item up a half step, 
    \item up a whole step, 
    \item down a half step, 
    \item down a whole step, and 
    \item no change.
\end{itemize}
After communicating individual instructions, the machine then provides a cue to all musicians to implement their instructed change. Figure \ref{fig:gestures} provides illustration of some of these robotic gestures.

\begin{figure}[h]
    \centering
    \includegraphics[width=\textwidth, trim={3cm, 4cm, 5cm, 2.5cm}, clip]{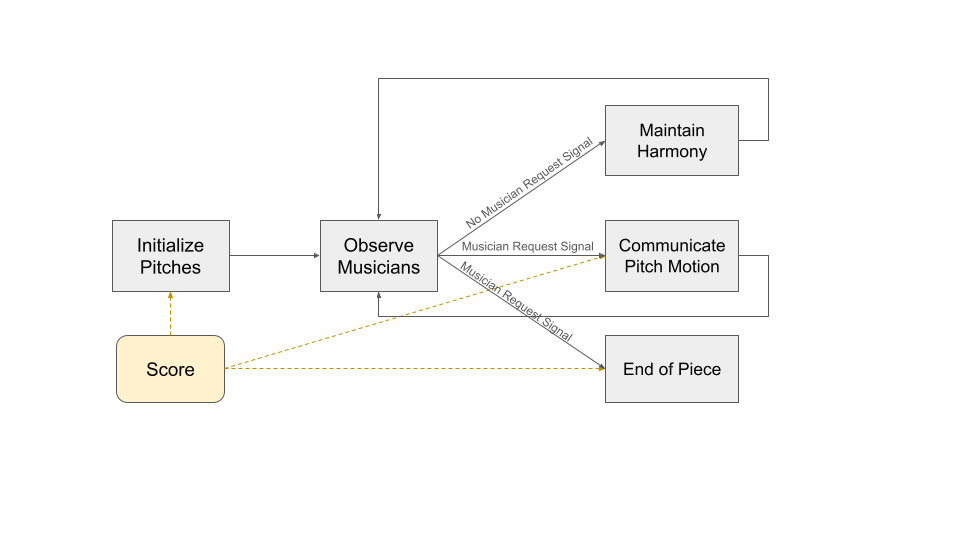}
    \caption{A high-level diagram of the ImproVision system. Based on a score, the machine initializes pitches for the musicians. The system observes the musicians, maintaining the harmony unless a Musician Request Signal (in this case, a nonverbal signal) is detected. If a Musician Request Signal is detected, the machine communicates the desired new pitches through movements. If the end of the score is reached, the machine conveys an end-of-piece signal. Note that the score is not necessarily fixed; it could merely be a generative reference throughout.}
    \label{fig:overview}
\end{figure}

\begin{figure}
    \centering
    \includegraphics[width=\textwidth]{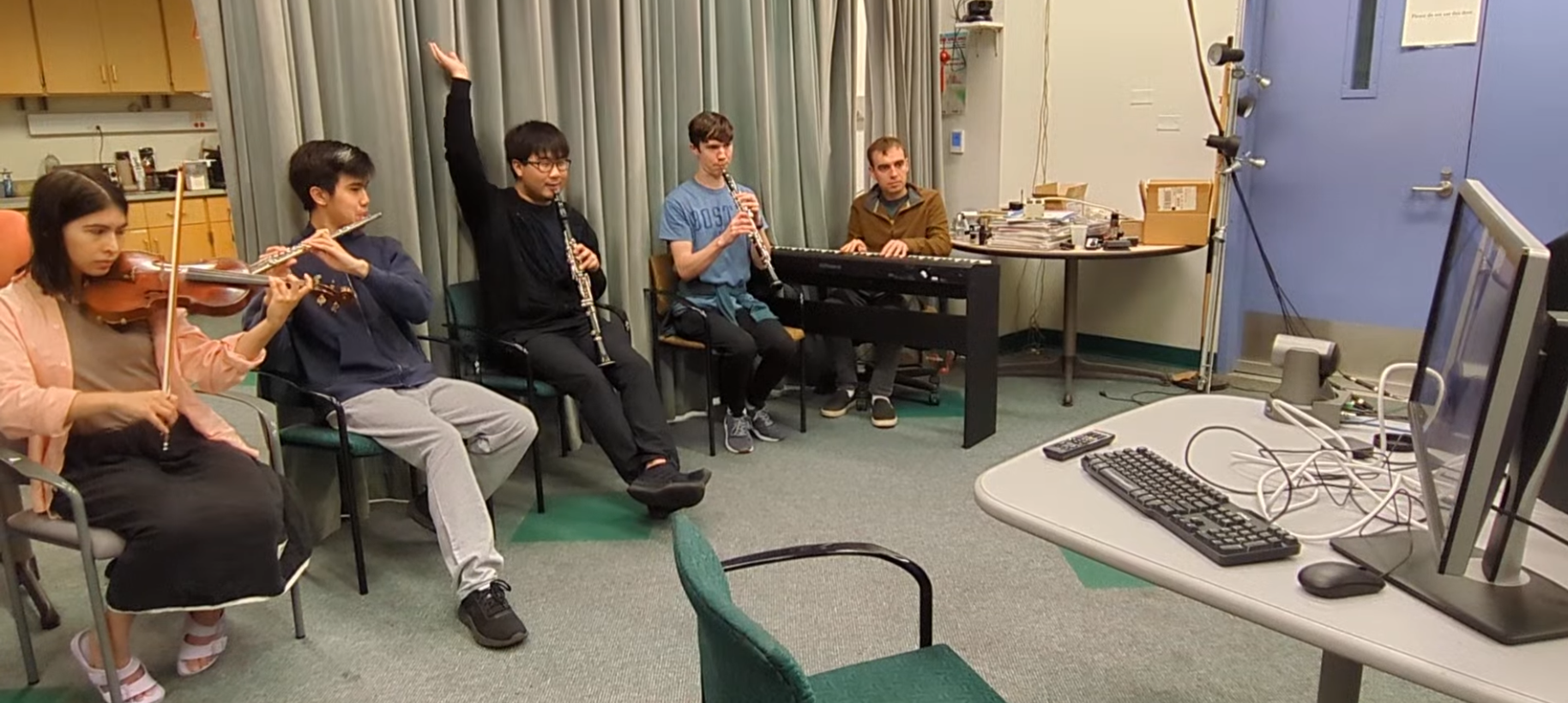}
    \caption{The clarinet player provides a signal to the ImproVision system that he is ready to proceed further in the score. The ImproVision system will observe the signal and then provide new instructions to the ensemble.}
    \label{fig:randy}
\end{figure}

\begin{figure}
    \centering
    \includegraphics[width=\textwidth]{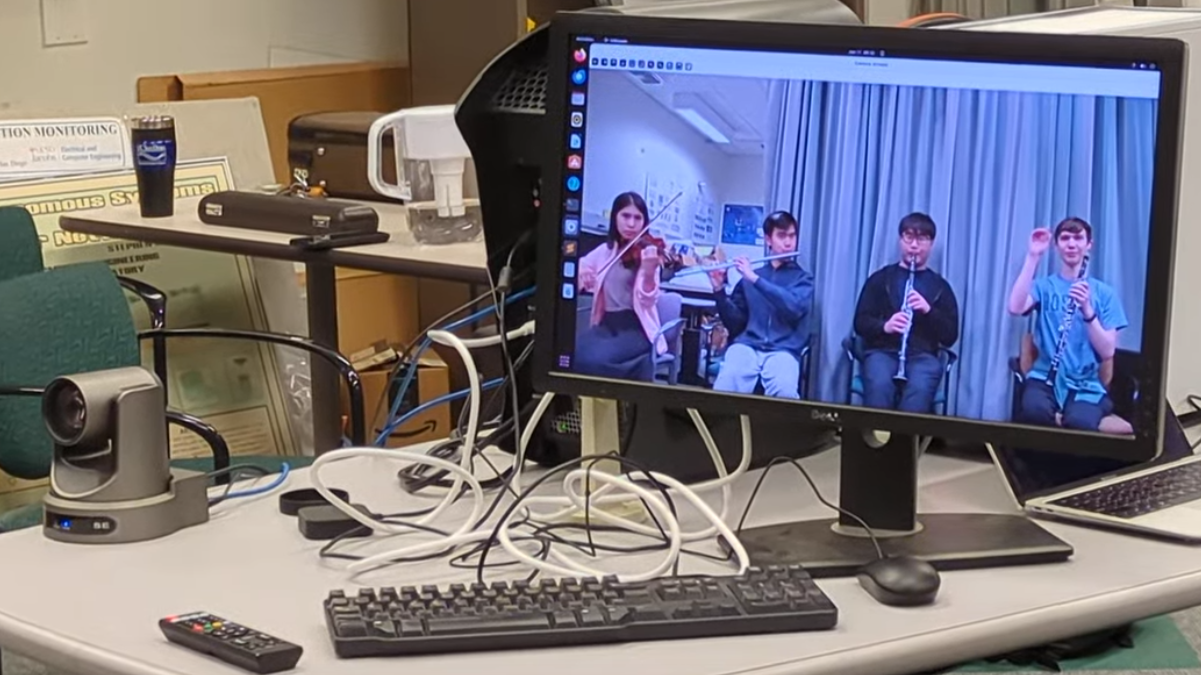}
    \caption{The ImproVision system maintains awareness of the state of each musician; in this setting, the system detects pose keypoints of the musicians in order to observe the predefined nonverbal cue (hand-raise).}
    \label{fig:computer}
\end{figure}

As a concrete example to illustrate the game more clearly, imagine a score with a C Major chord as measure one and an F Major chord as measure two. Corresponding to Measure 1, a violinist, violist, and cellist are initialized on the notes “C,” “E,” and “G,” respectively. The musicians sustain this chord until one of them, say the violist, expresses discontentment by raising her hand. (Note that the violist does not know anything about future measures, just that she no longer wants to stay on Measure 1.) The machine detects her raised hand. It knows that Measure 2 is an F Major chord, which means that the violinist should stay on C, the violist should move to F, and the cellist should move to A. First, since the violinist’s instruction is to stay on the same note without changing, the machine pans to the violinist, makes “eye contact” for a moment, and then pans to the violist. Second, the machine wants the viola to move from E up to F, so it gives the “up half step” cue. Third, the cellist receives the “up whole step” cue for moving from G to A. Finally, the machine faces the middle and provides a final cue to all musicians to implement their instructed change. The trio is now on an F chord. How long they sustain each individual chord is up to them. A diagram describing the system flow is provided in Figure \ref{fig:overview}. A short demonstration of a quartet playing this improvisational game is available online\footnote{https://github.com/rossgreer/ImproVision/}, with illustrative images in Figures \ref{fig:randy} and \ref{fig:computer}.

\section{In Support of Learning Methods for Co-creativity: ImproVision Extensions}

Due to the subjective nature of creativity, the level of expertise required of human agents, and difficulty in controlling for co-factors affecting human expression, there is a general lack of data supporting the learning of human-machine (and even human-human) interactions in co-creative artistic endeavors  \cite{dubnov2014delegating}. In this section, we describe possible extensions of our initial musical ``Guided Harmony" game, some data collection methods, and expected benefits and capabilities enabled by the collection of such data. We aim to make these descriptions specific enough to facilitate data collection and demonstration with sufficient detail, but presented in a way that allows for adaptation to more general application or variation. 

\subsection{Nonverbal Toolkit for Musicians}

Needless to say, using a raised hand as a cue is not an ideal choice. Raising one’s hand disrupts musical output, as many instruments require both hands for sustained sound. Accordingly, to provide a control signal to a machine improvisation tool, we identify the need to design a toolkit with the following properties based on our initial demo:
\begin{itemize}
    \item Signals must be nonverbal,
    \item Signals must be possible given the occupational requirements of the instrument (i.e., a pianist cannot signal with their hands, a flautist cannot signal with their mouth, a violinist cannot use their arms),
    \item Signals must be differentiable from natural gestures one might make while playing (e.g., blinking), i.e., the signal should not be too trivial,
    \item Signals should be coherent to the intended affect (e.g., human should not be closing their eyes at the most energetic, dance-like moment of a piece, nor lifting their eyebrows in surprise at a quiet, contemplative moment),
    \item Signals must be obvious enough for a camera to pick them up from several feet away in a busy scene.
\end{itemize}

It may become apparent that different instruments need different cue toolkits. In Table \ref{tab:sum}, we provide a non-exhaustive summary of identified communication avenues for both human and machine signals within the ImproVision framework and PTZ camera setup. While a PTZ camera does provide limitations to kinetic expression compared to robots with higher degrees of freedom, there is still a wealth of signals which can be generated for interaction.

\begin{table}[]
    \centering
    \caption{The Guided Harmony game, using a PTZ camera in the ImproVision framework, utilizes a variety of communication styles from this table. Machine audio output provides initial pitch settings. Human hand-raising and robotic eye contact, nodding, and panning provide information about who is expected to play,  what sounds they should contribute, and when they should be played.}
    \label{tab:sum}
\begin{tabular}{>{\centering\arraybackslash}p{1.2cm}|>{\centering\arraybackslash}p{5.2cm}|>{\centering\arraybackslash}p{5.2cm}}
     & \textbf{Human} & \textbf{Machine}  \\ \hline
    \textbf{Audio} & Instrument & Audio Output \\ \hline
    \textbf{Visual} & Raising Hand, Making Hand Gesture, Raising Foot, Movement of Body, Facial Expression, Pose, Eye Contact  & Looking Towards Human, Nodding Up, Nodding Down, Varying Speed of Motion, ``Panning" Rotation, ``Tilting" Motion \\
\end{tabular}
\end{table}

\subsection{Enhanced Focus on Score Generation and Creativity}

Rather than have the machine parse a human-composed score, it would be a logical extension of the existing system to give the machine agency in creating the score through integration with music generation models. Allowing the machine to recognize audio cues or, more generally, identify the chords being played would also make the system more robust against human error (i.e., if a musician misinterprets or forgets a cue, the system would be able to hear the mistake and adapt future cues to return to the intended chord progression).

We also imagine a scenario in which the machine's music generation algorithm attempts to infer the given musicians' chord preferences based on the time elapsed between a new chord being played and someone raising their hand. By targeting the optimization of group enjoyment, the creative process becomes more directed toward a clear objective, and may even reveal quantitative insights about the performers' (or population) musical aesthetic preferences. 

\subsection{Identification of the Improvisational Leader}

In cases when an ensemble is improvising together, there is a natural amount of turn-taking in allowing a soloist to lead ensemble play. If a machine agent can communicate, suggest, and direct such turn-taking, it enables the machine to be part of the musical flow, creating its own sequences of solos and, at times, directing action to support its own musical output. Toward this, with the use of a PTZ camera or other directional indicator, we propose demonstration of a system that actively listens to ensemble musical output and uses its communication channel to offer a suggestion (or direction) to the identity of the next lead player. 

\subsection{Silent Musical Charades}

Nonverbal communication of musical intent, pleasure, and frustration is difficult to capture, as many humans wear different physical expressions to communicate these ideas. Moreso, trained musicians tend to communicate these ideas in a multimodal setting, letting their sounded output co-communicate both the intended music and also their intent, coherence, or frustration with their stand partner. 

We also consider that the training of a musical improvisation system may benefit from a scalar reward function, to either provide a positive, encouraging reward for the system to continue on its path, or a negative, discouraging reward for the system to try something different. 

While the aforementioned toolbox provides one way for this to be communicated, it fails to be naturalistic as it is a pre-programmed codebook of control commands. Here, we propose a scheme for the collection of data which can represent the user's frustration and intent in a nonverbal manner.

\begin{itemize}
    \item Musician A and B begin with some shared repertory knowledge. For example, these may be musicians who had similar orchestral experience, played in the same jazz band, studied conservatory solo pieces on the same instrument, etc.
    \item Musician A is holding their instrument as if playing and imitates playing but will not create any sound.
    \item Musician A is given a selection of musical repertoire, and, in the style of charades, wins when Musician B guesses which piece was given. 
\end{itemize}

It is our hope that, in facilitating this demonstration, a few patterns may emerge:
\begin{itemize}
    \item Musician A will express ``pleasure" in a naturalistic, nonverbal way when Musician B provides a ``nearby" guess.
    \item Musician A will express ``frustration" in a naturalistic, nonverbal way when Musician B provides a ``distant" guess or fails to note some salient feature.
    \item Musician A will communicate in a way that is possible given the occupation of their instrument.
    \item Musician A will communicate salient features of the music in a non-verbal (albeit exaggerated) way, such that Musician B will understand not only the name of the resulting piece but also the associated musical features and playstyles \cite{argamon2010structure} \cite{cont2006framework} that contribute to their understanding of that piece. 
\end{itemize}

Accordingly, the critical data in this experiment is a strong view (or set of views) of the face and pose of Musician A during the interactions, as well as the record of the chosen piece and the guesses of Musician B to facilitate the annotation of their interactions with measurements of proximity of the guesses (or at least the demonstrated affect). From this data, a training corpus can be created, which represents an association between visual features of an actively-performing musician and degree of satisfaction or dissatisfaction with their partner. In dataset collection, this partner is a human, but from this data, interactions can be modeled for eventual replacement of the human partner with a robotic or computer partner that can infer and interpret the affect of the human musician, even while the human is physically occupied in performance posture. 

\section{Conclusion}

To conclude, the presented research and framework explores the intersection of visual communication and musical interaction by implementing a robotic camera in a ``Guided Harmony" musical game. By leveraging bidirectional communication between human musicians and a robotic system, we demonstrate the potential for co-creative human-machine interaction in a musical context. The game underscores the importance of robust and intuitive communication channels for enhancing the collaborative creative process.

A natural progression of this work would be refining the visual cue toolkit to accommodate the diverse needs of different musical instruments and settings, allowing the machine to participate actively by integrating music generation algorithms and closed-loop audio feedback, and collecting more data on human-machine interactions in these settings. Stated more generally, remaining challenges include
\begin{enumerate}
    \item defining and expanding the amount of signals or gestures conveyed during  performance,
    \item increasing the agency of the performers and the conductor,
    \item specifying the score in terms of partial information (graphic scores \cite{jvania2017cardew}),
    \item formalizing the computational language of writing the interaction (temporal logic, iscore, petri-nets \cite{peterson1977petri}), and 
    \item learning creative interaction patterns. 
\end{enumerate}

In further research, a basic difficulty is defining optimal co-creative interaction; possible definitions include maximizing the amount of influence that performers have on each other, creating the most informative joint result with least constraint on individual expression (least effort communications), etc. These interactions may be measured through directed information, transfer entropy \cite{dubnov2023switching}, communication inside the group, and information contents of the overall (joint) message. Ultimately, the presented research and derivative explorations contribute to the broader understanding of how AI can enhance human creativity in musical contexts.

\section{Acknowledgements}
This project is partially supported by the European Research Council under Europe’s Horizon 2020 program, grant \#883313 (ERC REACH). The authors would like to thank demo musicians Lucy Lennemann, Keene Cheung, Randy Lew, and Jacob Butler. 

%
%
%
\bibliographystyle{splncs04}
\bibliography{refs}

\begin{thebibliography}{10}
\providecommand{\url}[1]{\texttt{#1}}
\providecommand{\urlprefix}{URL }
\providecommand{\doi}[1]{https://doi.org/#1}

\bibitem{argamon2010structure}
Argamon, S., Burns, K., Dubnov, S.: The structure of style: Algorithmic approaches to understanding manner and meaning. Springer (2010)

\bibitem{boden2013creativity}
Boden, M.A.: Creativity as a neuroscientific mystery  (2013)

\bibitem{brackett2010some}
Brackett, J.: Some notes on john zorn’s cobra. American Music  \textbf{28}(1),  44--75 (2010)

\bibitem{cabibihan2012human}
Cabibihan, J.J., So, W.C., Pramanik, S.: Human-recognizable robotic gestures. IEEE Transactions on Autonomous Mental Development  \textbf{4}(4),  305--314 (2012)

\bibitem{cont2006framework}
Cont, A., Dubnov, S., Assayag, G.: A framework for anticipatory machine improvisation and style imitation. In: Anticipatory Behavior in Adaptive Learning Systems (ABiALS). ABIALS (2006)

\bibitem{davis2013human}
Davis, N.: Human-computer co-creativity: Blending human and computational creativity. In: Proceedings of the AAAI Conference on Artificial Intelligence and Interactive Digital Entertainment. vol.~9, pp. 9--12 (2013)

\bibitem{dubnov2023switching}
Dubnov, S., Gokul, V., Assayag, G.: Switching machine improvisation models by latent transfer entropy criteria. In: Physical Sciences Forum. vol.~5, p.~49. MDPI (2023)

\bibitem{dubnov2023deep}
Dubnov, S., Greer, R.: Deep and shallow: Machine learning in music and audio. CRC Press (2023)

\bibitem{dubnov2014delegating}
Dubnov, S., Surges, G.: Delegating creativity: Use of musical algorithms in machine listening and composition. Digital Da Vinci: Computers in Music pp. 127--158 (2014)

\bibitem{eco1989open}
Eco, U.: The open work. Harvard University Press (1989)

\bibitem{gordy1999duo}
Gordy, C.A.: The duo piano experience: nonverbal communication between ensemble musicians. Ph.D. thesis, Concordia University (1999)

\bibitem{greer2023safe}
Greer, R., Deo, N., Rangesh, A., Trivedi, M., Gunaratne, P.: Safe control transitions: Machine vision based observable readiness index and data-driven takeover time prediction. In: 27th International Technical Conference on the Enhanced Safety of Vehicles (ESV) National Highway Traffic Safety Administration. No. 23-0331 (2023)

\bibitem{greer2021restoring}
Greer, R., Dubnov, S.: Restoring eye contact to the virtual classroom with machine learning. arXiv preprint arXiv:2105.10047  (2021)

\bibitem{hedges1978dice}
Hedges, S.A.: Dice music in the eighteenth century. Music \& Letters  \textbf{59}(2),  180--187 (1978)

\bibitem{hoffman2010shimon}
Hoffman, G., Weinberg, G.: Shimon: an interactive improvisational robotic marimba player. In: CHI'10 Extended Abstracts on Human Factors in Computing Systems, pp. 3097--3102 (2010)

\bibitem{juslin1997emotional}
Juslin, P.N.: Emotional communication in music performance: A functionalist perspective and some data. Music perception  \textbf{14}(4),  383--418 (1997)

\bibitem{jvania2017cardew}
Jvania, N.: Cardew's treatise-graphic score. improvisation, interpretation, or composition? GESJ: Musicology and Cultural Science (2), ~16 (2017)

\bibitem{lin2015audio}
Lin, Y.T., Liu, I.T., Jang, J.S.R., Wu, J.L.: Audio musical dice game: A user-preference-aware medley generating system. ACM Transactions on Multimedia Computing, Communications, and Applications (TOMM)  \textbf{11}(4),  1--24 (2015)

\bibitem{peterson1977petri}
Peterson, J.L.: Petri nets. ACM Computing Surveys (CSUR)  \textbf{9}(3),  223--252 (1977)

\bibitem{rangesh2021autonomous}
Rangesh, A., Deo, N., Greer, R., Gunaratne, P., Trivedi, M.M.: Autonomous vehicles that alert humans to take-over controls: Modeling with real-world data. In: 2021 IEEE International Intelligent Transportation Systems Conference (ITSC). pp. 231--236. IEEE (2021)

\bibitem{van2013free}
Van~der Schyff, D.: The free improvisation game: Performing john zorn’s cobra. Journal of Research in Music Performance  \textbf{2013} (2013)

\bibitem{stanley2009butch}
Stanley, T.T.: Butch morris and the art of conduction  (2009)

\bibitem{Soundpainting}
Thompson, W.: Soundpainting. \url{https://www.soundpainting.com/soundpainting/}, accessed: 2024-06-29

\bibitem{volchenkov2012markov}
Volchenkov, D., Dawin, J.: Markov chain analysis of musical dice games. In: Chaos, Complexity And Transport, pp. 204--229. World Scientific (2012)

\end{thebibliography}
%




\end{document}